\documentclass[a4paper]{article}
\usepackage{cite}
\usepackage{wrapfig}
\usepackage{graphicx}
\usepackage{amssymb}
\usepackage{amsfonts}
\usepackage{amsmath}
\usepackage{epsfig}

\begin{document}

\title{The Equation of State\\ and Multiparticle Production }
\author{\large R.N.\,Rogalyov$^{a}$\footnote{{\it This is a preprint of the work  published  in "Physics of Particles and Nuclei", vol. 55} \copyright Pleiades Publishing,
Ltd. 2024 (http://pleiades.online)}\\[2mm] \small
{$^{a}$\,NRC ``Krchatov Institute'' - IHEP, Protvino, Moscow region, 142281 Russia}
}
\maketitle
\setcounter{footnote}{0}

\begin{abstract}
We discuss the distribution of fireballs
produced in heavy-ion collisions in the net-baryon number  
and argue that neither the Free-Quark Model (FQM)
nor the Hadron Resonance Gas (HRG) model
can provide a comprehensive explanation of the distribution 
observed at the LHC.
The concept of net-baryon number freezeout temperature is suggested
and the role of sea quarks as a possible source of net-baryon 
number fluctuations is emphasized.
\end{abstract}
\vspace*{6pt}


\section{Introduction}
\label{sec:intro}

For many years, the properties of strong-interacting matter produced in collisions of heavy nuclei have remained one of the most intiguing issues in high-energy physics.
Extremely hot and dense state of matter is approached in such collisions,
record values are obtained in lead--lead collisions at the LHC at the energy per nucleon--nucleon pair $\sqrt{s_{NN}}\approx 5$~TeV.

A brief outline of the time evolution of such a collision is as follows.
At the time $0.1\div 0.3$~fm$/c$ after collision, the energy
density approaches its peak and quarks and gluons produced in primary parton collisions form a fireball of quark-gluon matter, which then thermalizes 
to quark-gluon plasma (QGP) at the time $t \sim 1$~fm$/c$. 
Then the fireball further expands and cools to the temperature $T_f$
at which it decays into hadrons.
The temperature $T_f\approx 155$~MeV is referred to as the freezeout temperature,
$T_f\approx T_{pc}$, where $T_{pc}$ is the pseudocritical temperature of the chiral crossover.

The produced hadrons carry information about the last 
stage of QGP evolution, direct information on early stages can be 
extracted from a very few sources, for example, 
from direct photons and jets.
Of particular interest is the question of the dynamics of the
net-baryon number $B=N_B-N_{\bar B}$, where $N_B$($N_{\bar B}$) is the number of
baryons(antibaryons) produced in a paricular collision.
$B$ is a conserved quantity and, therefore, 
can also be used as a signal of an early stage of QGP evolution.
Though the baryon chemical potential $\mu$ of fireballs produced 
at the LHC energies vanishes and the corresponding 
average net-baryon number equals zero, $\langle B \rangle \approx 0$,
fluctuations of $B$ carry information on
the underlying dynamical mechanisms.

In this work, primary attention is concentrated on 
the relation between the equation of state of the fireball matter, its evolution,  
and the distribution of fireballs in the net-baryon number.

The cumulants of this distribution, measured by ALICE~\cite{Behera:2018wqk},
agree well with the predictions of the Hadron Resonance Gas (HRG) model
and disagree with the predictions of the Free Quark Model (FQM).
Therewith, the HRG predictions are related to the freezeout temperature
and do not properly take into account net-baryon number conservation.
Based on the analysis of dynamics of the the net-baryon density,
we argue that, in the case under consideration, 
even exellent formal agreement with the experimental data
cannot give a comprehensive explanation of 
the net-baryon density distributions within the HRG model.
In our qualitative speculations we restrict our attention to the 
version of HRG which is referred to as the ideal hadron resonance gas model.

In the following Section we consider the relation between
the Equation of State (EoS) derived either from lattice QCD 
simulations or from a QCD-inspired model
and the probability distribution in $B$.
In Section~\ref{sec:fve} we estimate 
finte-volume corrections and the effect 
of cutoff in transverse momenta in FQM
in order to understand whether they can 
decrease the disagreement with experimental data.
In Section~\ref{sec:probe} we discuss
in detail a comparison of the FQM and HRG predictions 
with experimental data as well as theoretical 
limitations of predictive power of these models.

\section{Equation of state and net-baryon number probability distributions}
\label{sec:density}

The grand canonical partition function
characterizing strong-interacting matter 
at a definite temperature $T$, volume $V$ and baryon chemical potential $\mu$
meets the general thermodynamical relations and, in particular, 
\begin{equation} \label{eq:ZGC_fugacity}
Z_{GC}(\theta)\;=\; \exp\left({p(\theta)V \over T} \right) \;=\; \sum_{B=-\infty}^\infty Z_C(B) e^{B\theta},
\end{equation}
where $\displaystyle \theta= {\mu\over T}\; $, $p(\theta)$ is the pressure, $Z_C(B)$ is the canonical partition function at the net-baryon number $B$; formula (\ref{eq:ZGC_fugacity})
is referred to as the fugacity expansion.
In what follows, we use dimensionless 
pressure $\displaystyle \hat p={p\over T^4}$ and the net-baryon number density
 $\displaystyle \hat \rho={\rho \over T^3}$, they 
are related by the formula $\displaystyle \hat \rho={\partial \hat p \over \partial \theta}$.
Given $\hat p$ and $\hat \rho$ as functions of $\theta$,
one readily obtains the EoS in the paramtric form.
At $T>T_{RW}$, where $T_{RW}$ is the Roberge-Weiss 
temperature~\cite{Roberge:1986mm}, it has the form~\cite{Bornyakov:2016ccq,Bornyakov:2022blw}
\begin{eqnarray}\label{eq:free_EoS_param}
{\rho \over T^3}&=&  a_1 \theta + a_3 \theta^3 \\ 
{p\over T^4} &=& {a_1\over 2} \theta^2 + {a_3\over 4} \theta^4  + \hat p_0\;.  \label{eq:free_press}
\end{eqnarray} 
In the infinite-volume limit of the FQM with zero triality~\cite{Faber:1995up}, one obtains $\displaystyle a_1={N_F\over 3N_C}$, $\displaystyle a_3={N_F\over 3\pi^2 N_C^3}$, this also corresponds to the results of simulations
of lattice QCD at $T>T_{RW}\sim200$~MeV~\cite{Bazavov:2017dus}.

At $T<T_{pc}$, the EoS can also be derived from the results of 
lattice QCD simulations~\cite{Bornyakov:2016ccq, Rogalyov:2023pym}, 
it is given by the relations
\begin{eqnarray}\label{eq:HRG_EoS}
\hspace*{13mm}{\rho \over T^3}&=&  f_1\; \mathrm{sh} \theta\;,  \\ \nonumber 
{p\over T^4} &=& f_1\; \Big(\mathrm{ch} \theta \ -\;1 \Big) + \hat p_0\,.  \nonumber 
\end{eqnarray} 
The same EoS can be derived from the HRG model~\cite{Braun-Munzinger:2011shf}
using the fact that the EoS of the fireball matter is connected with the distribution of the collision events in the net-baryon number in a definite rapidity window at midrapidity. Theoretically, the net-baryon number probability distribution is defined by the formula 
\begin{equation} Q_B(\theta)={Z_C(B)e^{B\theta} \over Z_{GC}(\theta)}
=P_B e^{B\theta}\;{ Z_{GC}(0)\over Z_{GC}(\theta)}\;,
\end{equation} 
where $P_B\equiv Q_B(\theta=0) $ is the net-baryon probability distribution at $\mu=0$, it involves all information on $\theta$-dependence.

From a probabilistic point of view, one can consider
the moment's generating functions
\begin{equation}
M_\theta(t) = {Z_{GC}(t+\theta)\over Z_{GC}(\theta) }  \qquad \mbox{and} \qquad  
\mathfrak{M}(t)={Z_{GC}(t)\over Z_{GC}(0) } 
\end{equation}
as well as the cumulant generating functions
\begin{equation}
 K_\theta(t) = \ln M_\theta(t) 
 \qquad \mbox{and} \qquad   
\mathfrak{K}(t)=\ln \mathfrak{M}(t) = {\big(\hat p(t) - \hat p(0)\big) \nu } 
\end{equation}
where $\mathfrak{M}(t)$ and $\mathfrak{K}(t)$ are related to the probabulity mass 
function $P_n$ and $\nu=VT^3$ is the dimensionless variable characterizing
the number of modes in volume $V$ excited at temperature $T$ ($\nu\sim (2\pi)^3 n_{modes}$).
The cumulants $\kappa_{2n}$ for the distribution in the net-baryon number and the respective generalized susceptibilities $\chi_{2n}$  are defined by the relation
\begin{equation}
\mathfrak{K}(t)= \sum_{n=1}^\infty \kappa_{2n} {t^{2n} \over (2n)!}\;,
\qquad \quad \chi_{2n} = {\kappa_{2n}\over \nu}  \;;
\end{equation}
only even cumulants differ from zero since $P_B=P_{-B}$
in view of $C-$parity conservation.
Since $\langle B^{2n+1} \rangle =0$, 
\begin{eqnarray}
\chi_2&=&{1\over \nu} \langle B^2\rangle\;, \nonumber \\ 
\chi_4&=&{1\over \nu} \left( \langle B^4\rangle - 3 \langle B^2\rangle^2 \right)\;, \nonumber \\
\chi_6&=&{1\over \nu} \left(  \langle B^6\rangle - 15 \langle B^2\rangle \langle B^4\rangle \right)  \nonumber
\end{eqnarray}
and so on.

In the FQM with $N_C$ colors and $N_F$ flavors,
the probability mass function associated with the 
pressure (\ref{eq:free_press}) has the form \cite{Bornyakov:2022blw}
\begin{eqnarray}
P_B &=& C(\nu) \exp\left\{ {\pi^2 N_F N_C \over 36} \left[ \big(q^{1/3} 
+q^{-1/3} \big) \sqrt{\nu^2+ {243 \, B^2\over 4\pi^2 N_F^2} } -2\nu \right] \right.  \\
&& \left. -\; {3 \pi N_C B\sqrt{3}\over 8} \big(q^{1/3} -q^{-1/3} \big) \right\}\;, \nonumber
\end{eqnarray}
where $C(\nu)$ is the normalization constant and
$$
q=\sqrt{ 1+ {243\over 4\pi^2 N_F^2} {B^2\over \nu^2}} + {9B\sqrt{3}\over 2\pi N_F \nu}\;.
$$
The leading term of asymptotic behavior at $B\to\infty (B\gg \nu) $
is given by the formula
\begin{equation}\label{eq:prob_freeq_asympt}
P_B \simeq C(\nu) \exp\left(\,-\; {3N_C \over 4 } \sqrt[3]{3\pi^2 B^4\over \nu N_F} \right) 
\end{equation}

The probability distribution of $B$ in the HRG model has the form
\cite{Braun-Munzinger:2011shf,Braun-Munzinger:2011xux}
\begin{equation}
P(B;N_b,N_{\bar b})=e^{-(N_b+N_{\bar b})} \left({N_b\over N_{\bar b}}\right)^{{B\over 2}} I_B(2\sqrt{N_b N_{\bar b}}),
\end{equation}
where
$\displaystyle N_b={\nu f_1\over 2}\, e^\theta$, $\displaystyle N_{\bar b}={\nu f_1\over 2}\,e^{-\;\theta}$ 
are the average baryon and antibaryon multiplicities
and $f_1\nu =2\sqrt{N_b N_{\bar b}}\;$, which is independent of $\mu$.
Asymptotic behavior at $B\to\infty (B\gg\nu)$  of the probability mass function at $\theta=0$ is given by the formula
\begin{equation}\label{eq:prob_HRG_asympt}
P(B)\simeq \exp\left[ -B \ln \left({2B\over f_1 e}\right) - {1\over 2} \ln (2\pi B) -f_1\right]\;.
\end{equation}
\newpage
\section{Finite-volume and acceptance effects}
\label{sec:fve}

From formulas (\ref{eq:prob_freeq_asympt}) 
and (\ref{eq:prob_HRG_asympt}) it follows that
high-$B$ fluctuations in the FQM 
are substantially suppressed as compared with the HRG model.
This comparison can be quantified by 
considering the cumulant ratios.
In the FQM we obtain
\begin{equation}\label{eq:cum_ratio_FQM}
{\chi_4 \over \chi_2}={1\over 2\pi^2 N_c^2}\;; \qquad {\chi_n \over \chi_2}=0\quad \mbox{at}\quad n>2\;, 
\end{equation}
whereas in the HRG model we have\cite{Karsch:2010ck}
\begin{equation}\label{eq:cum_ratio_HRG}
{\chi_{2n} \over \chi_2}=1\quad \forall n\;. 
\end{equation}

However, it should be noticed that the final-volume corrections to 
the FQM can in principle change theoretical predictions of the cumulants, 
so we perform a rough estimate of such effects
assuming that the net-baryon number 
in the rapidity bin under study  ceases to fluctuate
at the moment of time $\tau\sim 2$~fm/$c$, when the baryon number 
at midrapidity ceases to change. We can divide 
the fireball --- the volume between two disks of radius $R=7$~fm
and distance between them $c\tau\sim 2$~fm ---
into $\sim 40$ cubes with side of length $c\tau$ ($c$ is the speed of light)
so that physical processes in each cube can be considered as independent 
of the others. So we estimate the distribution 
in the net-baryon number for an ensemble of such cubes
considering that each cube has a contact with the thermostat at $T\sim 300$~MeV 
and $\mu=0$, thus $\nu\sim 30$. 

We perform direct computations of the partition function
of the zero-triality gas of massless free spin-1/2 quarks in a cube
of volume $V$ by the formulas
\begin{equation}\label{eq:ZGC_compu1}
 Z_{GC}(\theta)= {1\over 3} \left[\tilde Z_{GC}(\theta) + \tilde Z_{GC}\left(\theta-\;{2\imath \pi \over 3}\right) +  \tilde Z_{GC}\left(\theta + {2\imath \pi\over 3 } \right)  \right]\;,
\end{equation}
where
\begin{equation}\label{eq:ZGC_compu2}
\! \tilde Z_{GC}(\theta)\! = \!
\prod_{\mathbf{n}\in \mathbb{Z}^3 } \left\{
\left[1+ \exp\left(\theta - {2\pi |\mathbf{n}| \over \sqrt[3]{\nu}} \right)\right]\!
\left[1+ \exp\left(-\theta - {2\pi |\mathbf{n}| \over \sqrt[3]{\nu}} \right)\right]\right\}^{2 N_c N_F }
\end{equation}
and $\mathbf n=(n_1,n_2,n_3)$, where $n_i$ are integers.

\begin{table}[hhh]
\begin{center}
\begin{tabular}{|c|c|c|c|} \hline
  Specification of the FQM        &  $\kappa_2$   & $\kappa_4/\kappa_2$ & $\kappa_6/\kappa_2$ \\ \hline\hline
$\nu=30$, direct computation  &  6.68(2) & 0.0657(2)    &  0.00139(2)  \\
$\nu=30$, $p_\perp>300$~MeV   &  3.62(1) & 0.0857(2)    &  0.00202(3)  \\   \hline 
$\nu=30$, formula (\ref{eq:free_press}) &  6.667   & 0.0056       &  0.00000      \\
\hline\hline
\end{tabular}
\caption{ Dispersion $\kappa_2$ and the cumulant ratios in the FQM with $N_F=2, N_C=3,$ spin 1/2, and zero triality. 1st row: the results of computation by formulas (\ref{eq:ZGC_compu1}) and (\ref{eq:ZGC_compu2}) taking into account all momenta;  2nd row: the same, but with the momentum cutoff. The standard values are given in the 3rd row for comparison.}
\label{tab:FQM_cuts}
\end{center}
\end{table}
The results are presented in the table. 
The values obtained by the textbook formula (\ref{eq:free_press})
with the neglect of the finite-volume and acceptance effects
for the pressure are shown in the 3rd row of the table for comparison. 
Though these effects increase the cumulant ratio 
$\displaystyle {\kappa_4\over \kappa_2}$ 
by an order of magnitude and make 
$\displaystyle {\kappa_6\over \kappa_2}$
nonzero, a dramatic difference between 
the predictions of the FQM and HRG models remains.

\section{Net-baryon number fluctuations as a probe of high-T QGP}
\label{sec:probe}

Currently, the main attention in the literature is focused on
fluctuations of the net-baryon number in a narrow 
rapidity interval around the midrapidity.
It is generally considered that these fluctuations 
are related to the freezeout temperature
$T_f\lesssim T_{pc}$, where the Hadron Resonance Gas (HRG) model works well.

Some deviations of the HRG predictions from  
experimental data for cumulant ratios are explained by 
the so called corrections for the conservation 
of the total net-baryon number~\cite{Bzdak:2012an}. 
Such corrections were computed under the assumption that
there exists a domain of phase space 
(say, the full phase space) in which 
{\it (i)} the total net-baryon number is conserved 
in the process of fireball evolution 
and {\it (ii)} both $N_B$ and $N_{\bar B}$ follow
a Poisson distribution both in a narrow cell of this domain at midrapidity
(where experimental measurements are performed)
and in the complement of this cell. 

To study the effects of conservation
of the net-baryon number in more detail, 
one should first consider its distributions in rapidity.

The net-baryon number rapidity distribution
differs substantially from the rapidity distributions 
of the entropy, transverse energy or particle 
multiplicity \cite{Du:2022yok,Li:2023kja}.
At the energies $20 <\sqrt{s_{NN}}<200$~GeV, 
it has two peaks at certain forward and backward rapidities,
which become more pronounced and move away from each other
with an increase of the beam energy.
These peaks are conventionally interpreted as the 
partially stopped baryons \cite{Videbaek:2009zy}.
The whole distribution can be considered 
as the sum of three components: the two peaks 
and the plateau, the contribution of the plateau
decreases with the rise of the beam energy.

Therefore, experimentally measured rapidity distributions
of the net-baryon density indicate that, 
at the beam energies below $200$~GeV, baryons 
from colliding nuclei find their way to all rapidity intervals.
For this reason, one cannot isolate 
rapidity interval other than the full rapidity range
in which the net-baryon number is conserved.

A completely different situation occurs at higher energies.
At the energies $\sqrt{s_{NN}}\gtrsim 5$~TeV 
one can consider the domain $D$ comprising
the rapidity range $|y|\lesssim y_c\approx 4$, where 
the average net-baryon number vanishes
and so the baryons from colliding nuclei do not penetrate to 
this domain. In the color glass condensate 
model of the early-stage evolution it turns zero at time 
$\tau\sim 2$~fm/c after collision \cite{McLerran:2018avb}.
At later moments of time, the flux of baryons into/from 
the domain $D$
ceases and then the net-baryon number is conserved in time;
its value is fixed at time $\tau$.
The temperature $T_{Bf}$ of the fireball at this moment
can be considered as the net-baryon-number freezout 
temperature, it is significantly higher than $T_{pc}$. 
Fireball at this time should be considered as 
strong-interacting matter at $T\sim T_{Bf}$,
and $\mu=0$ in the volume $V=\pi R^2 \tau$, 
where $R$ is the radius of colliding nuclei.
That is, the fireball before net-baryon freezout 
is described in the grand canonical approach, and after---
in the canonical approach (net-baryon number is conserved),
so that the net-baryon number fluctuctions 
in the domian $D$ are determined by the 
grand canonical partition function
$Z_{GC}(\theta,T_{Bf},V)$.

Thus at $T\sim T_{Bf}$, the 
net-baryon number density and the respective contribution to the pressure 
can be described by formula (\ref{eq:free_EoS_param}) 
with the parameters 
$\displaystyle a_1={N_F\over 3 N_C}, \ a_3=  {N_F\over 3 \pi^2 N_C^3}$
characterizing an ensemble of massless non-interacting fermions with $N_F$ flavors and $N_C$ colors at zero triality~\cite{Faber:1995up},
the finite-volume and acceptance effects 
can be neglected in our reasoning.
This statement is substantiated by 
lattice QCD studies~\cite{Bazavov:2017dus} indicating that, at the temperatures 
$T>200\div 230$~MeV, the cumulant ratios $\displaystyle {\kappa_4\over \kappa_2}$ 
and $\displaystyle {\kappa_6\over \kappa_2}$ associated with the 
probability distribution in the net-baryon number 
coincide with those for the FQM. 

At sufficiently high beam energy $\sqrt{s_{NN}}$,
net-baryon number conservation during the fireball 
evolution at times $t>\tau$ over the rapidity
window $|y|<y_c$ implies that 
the observed fluctuations in
the respective acceptance are related to
the temperature $T_{Bf}>T_{pc}$ rather than to
$T_f\lesssim T_{pc}$. From the above reasoning
it follows that the cumulant ratios at 
$\sqrt{s_{NN}}\sim5$~TeV should differ significantly from those
at $\sqrt{s_{NN}}<200$~GeV.

However, experimental data obtained by ALICE~\cite{Behera:2018wqk}
give cumulant ratios that agree with the HRG model 
and disagree with the FQM.

Nevertheless, the HRG model can hardly be applied to 
the case of extremely high collision energies
for the following reason.
The HRG prediction of large fluctuations of $B$ is related 
to the case when the system has a given value of chemical 
potential and $B$ is not conserved,
that is, $\mu$ describes contact of a system with the
external source of baryons and atibaryons. 
In the case under consideration, ``the system'' is 
a part of the phase space related to the acceptance
of an experiment and the ``thermostat'' is the remaining 
part of the phase space.
At TeV beam energies, the exchange of net-baryon number between them 
at the final stage of the fireball evolution
can be neglected: otherwise the diffusion of baryons 
in rapidity space leads to the penetration of baryons
from the baryon-rich domains at high rapidity
to the acceptance around midtapidity
giving rise to nonvanising values of $\mu$
in contrast to the experimental values $\mu<1$~MeV~\cite{ALICE:2023ulv}.

Therefore, the distributions of the net-baryon number in rapidity 
as well as the experimental values of $\mu$ imply that, say,
 a significant excess of antibaryons over baryons
cannot be produced with high probabilities at midrapidity
and $\mu\approx 0$ at the final stage of the fireball evolution,
where the HRG model is applicable.
On the other hand, the observed cumulant ratios of
the distribution in the net-baryon number
give evidence for non-negligible probability of
negative values of $B$.
Thus we conclude that, in spite of qualitative agreement
with the experimental data, the HRG model cannot explain
the values of net-baryon number fluctuations observed at the LHC.

Yet another point in favor of this conclusion is as follows.
Initial temperature of the fireball at the collision energy
$\sqrt{s_{NN}}\sim 5$~TeV is $T\approx 300$~MeV,
thus the subsequnt evolution of the fireball 
proceeds in the temperature range $200<T<300$~MeV,
where the FQM works well (according to lattice estimates)
and this stage should be characterized by the 
probability distribution (\ref{eq:prob_freeq_asympt})
with the cumulant ratios (\ref{eq:cum_ratio_FQM})
rather than (\ref{eq:prob_HRG_asympt}) with the cumulant ratios 
(\ref{eq:cum_ratio_HRG}).
At the temperatures below $200$~MeV
a rearrangement from the distribution
(\ref{eq:prob_freeq_asympt}) to the substantially different distribution (\ref{eq:prob_HRG_asympt})
is impossible in view of the net-baryon number conservation.

From this it is inferred that the observed 
distribution in $B$ is formed at the initial stage of collision
before thermalization.
Since it represents the Skellam distribution,
a hypothesis of some mechanism of independent production of 
quarks and antiquarks in the relevant rapidity window 
at an early stage of evolution is plausible.

In any case, since the predictions of FQM is inconsistent with the experimental data and the HRG model does not explain sufficiently rapid
penetration of the net-baryon number to the acceptance at midrapidity, the  distribution of fireballs in $B$ observed at the LHC is yet to be explained.

A possible explanation may be provided 
by considering the contributions of sea quarks
of colliding nuclei to the net-baryon number in
the relevant rapidity window 
at an early stage of evolution before thermalization
because the contributions of quarks and antiquarks
in this case are independent of each other.
However, a detailed analysis of this possibility
should be the subject of another study.

\label{sec:finish}
\section{Conclusions}

From the analysis of the net-baryon number distributions
in rapidity and the evaluations of the baryon chemical potential 
of fireballs produced at $\sqrt{s_{NN}}\sim 5$~TeV
it has been concluded that, at the LHC energies, 
the transport of net-baryon number in rapidity space is strongly suppressed at midrapidity.
This conclusion has been used for the analysis 
of the FQM and HRG predictions of the distribution
of the fireballs in $B$ at midrapidity
given the EoS in these models agree well
with the results obtained in lattice QCD.

As the temperature of the fireball decreases
from 300 to 200~MeV, the EoS of 
strong-interacting matter in the fireball
corresponds to the FQM distribution
provided that the net-baryon number freezeout 
does not occur at this stage.
The FQM distribution differs substantially from the HRG distribution
at great net-baryon numbers and
thus the former cannot be 
rapidly rearranged to the latter due to 
the suppression of the net-baryon transport.
 
For this reason, the HRG model is not applicable to 
the net-baryon number dynamics at the final stage of the fireball evolution.
The observed distribution of fireballs in $B$
should be attributed to an early stage of the 
fireball evolution, where net-baryon number freezeout occurs 
at $T>T_{pc}$.
This suggests to introduce the net-baryon number freezeout temperature $T_{Bf}\gtrsim 300$~MeV. In this case, fluctuations of the net-baryon number 
can arise due to the penetration of sea quarks of colliding nuclei 
into the region around midrapidity at an early stage of the collision.

\vspace*{5mm}

{\bf Acknowledgment}\\
This work was financed from the budget of the Federal State Budgetary Institution "Institute of High Energy Physics" of the National Research Center ``Kurchatov Institute''. No additional grants were received to conduct or direct this specific study.


\end{document}